\documentclass[12pt]{article}
\usepackage{amsmath}
\usepackage{graphicx}
\usepackage{color}
\usepackage{varioref}
\usepackage [margin=1in, paperwidth=8.5in, paperheight=11in]{geometry}

\begin{document}
\title\centerline{\Large{\textbf
{Role of Wake Potential on Self-diffusion of Dust
\newline \centerline{  Particles in Dissipative System}}}}
\begin{center}{Mahmuda Begum$^{\rm 1a}$  and  Nilakshi Das$^{\rm 1}$}\end{center}
\begin{center}{{$^{1}$\emph{Plasma Physics Research Laboratory, Department of Physics, Tezpur University,
 Tezpur, Napaam, Assam-784028, India}}}\end{center}
\begin{center}$^{a}${Electronic mail: mbegumtu@gmail.com
                                      }\end{center}
\begin{abstract}
 In this paper, the structural property of complex plasma and self-diffusion coefficient of dust particles in presence of such wake potential have been investigated using Langevin dynamics simulation in the subsonic regime of ion flow. The study reveals that the self-diffusion of dust grains is strongly affected by ion flow and it changes its character in the wake potential dominant to the Debye-H$\ddot{u}$ckel potential dominant regimes. The dependence of self-diffusion coefficient on parameters such as neutral pressure, dust size, ion flow velocity, Coulomb coupling parameter have been calculated for subsonic regime using Green-Kubo expression which is based on integrated velocity autocorrelation function (VACF).
\end{abstract}

\begin{paragraph}
{\textrm{I.}}\textbf{INTRODUCTION}

The charged dust particles immersed in plasma usually interact via isotropic Debye-H$\ddot{u}$ckel potential. Spherical 3D plasma crystals [1] with a nested shell structure have been created in bulk plasma of RF discharge
where plasma flow is negligible. In presence of ion flow, the interaction among the dust grains in plasma is multiscale in nature with a mixture of short range repulsive and long range attractive forces and this leads to emergence of many new effects. In such case, it is expected that self-diffusion in presence of potential will exhibit interesting behaviour.

Vertically aligned structures are observed in number of complex plasma experiments [2,3] in recent past. The alignment of dust particles has been explained on the basis of attractive wake potential that arises due to the focussing of
flowing ions by the negatively charged dust grains. The overshielding of the dust grains due to excessive accumulation of positive ions leads to the deformation of spherical debye sphere to an elongated wake. The result is that dust
particles are effectively trapped under an oscillatory attractive wake potential. The nature of this wake potential has been studied by number of authors like Ishihara et al. [2], Lemon et al. [4], Nambu et al. [5], Saurav et al. [6] etc. The role of wake potential in the formation of dust Coulomb crystals in a plasma with a finite ion flow was
experimentally demonstrated by Takahashi et al. [3]. They showed that a dust grain located in the up stream region of
ion flows caused an attractive force on another dust grain in the lower part and the grains undergo vertical alignment along the direction of ion flow. Very recently Goree et al. [7] investigated experimentally the ion flow effect on velocity correlations in dusty plasma .

The present study is aimed to investigate the structural properties and diffusion coefficient of dust grains along the vertical i.e. direction of ion flow where wake potential plays dominant role for certain values of ion flow velocities and along horizontal direction where Debye-H$\ddot{u}$ckel potential is effective. The understanding of transport
properties of complex plasma in presence of combined effect of wake and Debye-H$\ddot{u}$ckel potential may be
significant from the point of view of electrorheological properties of complex plasma.
Our simulation is based on the Langevin dynamics method which brings the simulation closer to real dusty-plasma experiments, as the effects of the neutral gas damping and dust particle Brownian motions are included in a self-consistent manner.

\end{paragraph}

\begin{paragraph} {\textrm{II.}}\textbf{THEORY AND SIMULATION}

We consider a 3D dusty plasma composed of electrons, ions, neutral particles and micron sized negatively charged dust particles. The dust particles, which are usually levitated in the plasma sheath encounter a downstream ion flow with uniform velocity along z-direction towards the plasma sheath.

   The dust-dust interaction between a pair of $i^{th}$ and $j^{th}$ dust particles may be represented by repulsive  Debye-H$\ddot{u}$ckel (DH) potential in general as [8,9]

\begin{equation}
\Phi_{Y}(r_{ij})=\frac{Q_{d}^{2}}{4\pi\epsilon_{0}r_{ij}}\exp\left(-\frac{r_{ij}}{\lambda_{D}}\right)\end{equation}
where	$Q_{d}$  is the particle charge, $\lambda_{D}$ is the screening length due to electrons and ions. $r_{ij}$
is the distance between two particles.

 In presence of streaming ions, a wake potential arises, which may be written as [6]

 \begin{equation}
\Phi_{W}(0,z)=\frac{Q_{d}^{2}}{4\pi\epsilon_{0}\lambda_{De}}2\left(\frac{2\pi}{6zM}\right)^{1/2}sin\left(\frac{z}{M}\right)
\end{equation}
Here, $M$=$\frac{u_{i0}}{c_{i}}$ is the mach number having $c_{i}= \lambda_{De}\omega_{pi}$ as ion acoustic wave
velocity, $u_{i0}$ is the equilibrium ion streaming velocity and $\omega_{pi}$ beings the ion plasma frequency. This potential is attractive, oscillatory in nature and effective only in ion flow direction. It is basically a function of the Mach number which represents ion flow velocity normalized with respect to dust-ion acoustic velocity. Thus, in presence of ion flow, the inter-particle interaction in plasma and hence the entire thermodynamics of such a system is controlled by three basic parameters-the Coulomb Coupling parameter $\Gamma$=$\frac{Z^{2}e^{2}}{4\pi\epsilon_{0}aK_{B}T_{d}}$, the screening parameter $\kappa$=$\frac{a}{\lambda_{D}}$ and the Mach number $M$,
where where `Z' is the total number of electrons accumulated on the dust particle and  $Ze=Q_{d}$, is the charge of a dust particle, `e' is the charge of an electron and $k_{B}T_{d}$ is the kinetic energy of dust particles. $k_{B}$ is the Boltzmann constant, `a' is the typical mean
 inter-particle distance denoted as the Wigner-Seitz radius, $n_{d }$ is the dust number density of dust particles,
 $T_{d}$ is the temperature of the dust grains. The system is called “strongly coupled” if the Coulomb coupling
 parameter  $\Gamma$ i.e., the ratio of the average inter particle potential energy to the average kinetic energy, is
 comparable with or greater than unity.

The aim of the present work is to examine the structural and diffusion property of the dusty plasma which is described by combined  Debye-H$\ddot{u}$ckel and wake potential.

\end{paragraph}

%\begin{paragraph} {\textbf{ Langevin MD Simulation:}}
\begin{paragraph} {\textbf{Langevin dynamics simulation:}}

  The numerical simulation performed for the present analysis is based on Langevin MD Simulation where the effect of dissipation due to neutral pressure is self-consistently taken into consideration. Langevin equation of motion for the $i^{th}$ dust particle may be written as [10,11]

\begin{equation}\frac{dv}{dt}=-\gamma v+\frac{F}{m_{d}}+A(t)\end{equation}
 where position and time are normalized by dust Debye length $\lambda_{D}$ and
 $\sqrt{\frac{m_{d}\lambda_{D}^{2}}{k_{B}T_{d}}}$ respectively.

 This equation includes frictional drag $-\gamma v$, random force $A(t)$ and the deterministic force $F$. In our system, along z-direction the interaction force F have been derived from the equation: $F(t)=-Q_{d}\sum\nabla_{i}\Phi$ having $\Phi=\Phi_{Y}+\Phi_{W}$, where $\Phi_{Y}$ is the Debye-Hückel potential given by equation (1) and $\Phi_{W}$ is the wake potential along the direction of ion flow given by equation (2).
On the plane perpendicular to the direction of ion flow the interaction force F have been derived from the equation: $F(t)=-Q_{d}\sum\nabla_{i}\Phi_{Y}$, where $\Phi_{Y}$ is the Debye-Hückel potential.

The term $-\gamma v$ represents dynamical friction of dust grains with background neutral particles and in normalized
form it is given as
\begin{equation}\gamma v=\left(\frac{4\pi}{3}\right)N_{n}\frac{m}{m_{d}}r_{d}^{2}cv\delta\end{equation}
where $\gamma =\left(\frac{4\pi}{3}\right)N_{n}\frac{m}{m_{d}}r_{d}^{2}c$, is the drag
coefficient. $N_{n}$, $m$, $r_{d}$, v and $c$ represents the normalized neutral density, neutral mass, dust radius, dust velocity
and neutral thermal velocity respectively. $\delta$ is the numerical constant for the Epstein drag force having
value $\delta\approx1.48$.

The term $A(t)$ represents the Brownian acceleration which is related to the random impulse exerted on dust grains by
the background neutral pressure due to frequent collision with dust grains. The effect of these collision can be
modeled by Gaussian white noise and in normalised form it is written as
\begin{equation}A(t)=\sqrt{\frac{\beta^{2}}{dt}}N(0,1)\end{equation}
where N(0,1) represents the normal random variable with mean 0 and variance 1. $\beta$ is an undetermined coefficient
and in equilibrium it is related to the drag coefficient $\gamma$ through fluctuation-dissipation theorem as
$\beta^{2}=2\gamma$.

The simulation is performed for 500 particles in a 3D cubic simulation box of side L. Velocity-verlet algorithm is used to calculate the new position and velocities from the computed Langevin equation of motion. The space, mass, time,
velocity and energy are normalized by $\lambda_{D}$, $m_{d}$, $\sqrt{\frac{m_{d}\lambda_{D}^{2}}{k_{B}T_{d}}}$, $\sqrt{\frac{m_{d}}{k_{B}T_{d}}}$ and $k_{B}T_{d}$.

Recently, the effect of ion-neutral collision on the formation of ion-flow induced wake potential has been studied [12,13,14]. It is found that in strong collisional limit, the ion focusing may be significantly reduced which results in deformation of wake potential. Maiorov et al. [15] have also reported that the vertical linear structure arising due to ion flow is suppressed due to ion-neutral collision.
However, in the present analysis the effect of ion-neutral collision has not been taken into account. Our simulation has been carried out in a regime where ion-neutral collisional mean free  path ($\sim10^{4}mm$ for neutral density $N_{n}=10^{18}m^{-3}$ and ion neutral collision cross section $\sigma_{in}\sim5.0\times10^{-19}m^{2}$) exceeds Debye length $\lambda_{De}$, which of the order of $10^{-2}m$ for the parameters chosen here. Collisionless treatment is therefore valid in this regime.

The information about structural properties of the system are obtained by calculating the radial distribution function
$g(r)$ and static structure factor $S(k)$ [16]. In a 3D system the pair correlation function is defined as
\begin{equation}g(r)=\frac{V}{N}\frac{N(r, \Delta)}{4\pi r^2\Delta}\end{equation} where $V$ is the volume of the simulated region, $N$ is the
number of simulated particles, and $N(r, \Delta)$ is the number of particles located in a shell of infinitesimal
thickness $\Delta$ from $r-\frac{\Delta}{2}$ and $r+\frac{\Delta}{2}$.
The static structure factor is defined as [16]
 \begin{equation}S(k)=1+\frac{N}{V}\int g(r)\exp (-ik.r)dr\end{equation}
  We choose for FCC lattice $k=(2\pi/L)(1,-1,1)$, where L is the length of the cubical simulation box.

The self-diffusion coefficient D of a system of particles in three dimensional case can be calculated using velocity
autocorrelation function (VACF) through the Green-Kubo integral formula [16]
\begin{equation} D=\frac{1}{3}\int_0^\infty Z(t)dt \end{equation}
where \begin{equation} VACF=Z(t)=\langle v_{j}(t)\cdot v_{j}(0)\rangle \end{equation}
The brackets $\langle...\rangle$  in equation (9) represent the canonical ensemble average over all particles. The
integrand Z(t) is the velocity autocorrelation function, which is calculated over all segments of the ensemble average
of the velocity products at time t and an initial time $t_{0}$.
\end{paragraph}
\begin{paragraph}{\textrm{III.}}\textbf{RESULTS AND DISCUSSION}

In this paper we have investigated the structural property and self-diffusion of dusty plasma in presence of streaming ion flow in the subsonic regime using Brownian dynamics simulation. In order to see the effect of neutral pressure on the motion of the dust particles, we have plotted self-diffusion coefficient across neutral pressure for different Mach numbers in Figure 1. It is found that self-diffusion coefficient decreases with the increase in pressure.

 The size of the grains plays a crucial role in the transport of complex plasma. We have carried out our simulation for particles with different size and the results are plotted in Figure 2. It is observed that with the rise in size of the grains, self-diffusion coefficient gradually decreases.

The main purpose of the present study is to see the effect of ion flow on the structure and diffusion of dust particles. Radial distribution function g(r) and static structure factor S(k) are plotted in Figure 3 and 4 for three different regimes of ion velocity corresponding to M=0.1, 0.3 and 0.5 respectively. Both these studies reveal that the peak height increases with the increase in Mach number. On the other hand, the effect of ion flow velocity on  self-diffusion coefficient is depicted in Figure 5. Here, self-diffusion coefficient is plotted across Coulomb Coupling parameter $\Gamma$ for four different values of Mach numbers M=0.1, 0.2, 0.3 and 0.5 respectively. This study reveals that self-diffusion coefficient gradually decreases with the rise in the value of Coulomb Coupling parameter $\Gamma$, indicating that the diffusion of the particles decreases when they are in more ordered state corresponding to the higher values of Coulomb Coupling parameter $\Gamma$ at fixed values of screening parameter $\kappa$. It is also seen that with the rise in the value of Mach number, i.e. normalized ion flow velocity, the system goes to more ordered state in the subsonic regime for the range of parameters studied in Figures (3, 4, 5) and the self-diffusion coefficient gradually decreases. It is clear from equation (2) that with the decrease in Mach number, the strength of wake potential increases. The particles experience a pull along the vertical direction and this effectively reduces the repulsive Yukawa potential. The result is that the particles go to a disordered state and the diffusion also increases. In Figure 6, we have plotted parallel component of self-diffusion coefficient across $\Gamma$ for various values of Mach number in the subsonic regime and these results also support above physical effects.

   In order to identify the dominant interaction among the particles for a particular value of mach number, we evaluate spring constant associated with the system. The spring constants coupled with Debye-H$\ddot{u}$ckel and wake potential may be defined  respectively as

$K_{Y}$=$\frac{d^{2}\Phi_{Y}}{dr^{2}}|_{r=a}$=$\frac{\Gamma\exp(-\kappa)}{\kappa^{2}a^{2}}\left[1+\kappa+\kappa^{2}\right]$

and $K_{W}$=$\frac{d^{2}\Phi_{W}}{dz^{2}}|_{z=a}$=$-2\Gamma\sqrt{\frac{2\pi\kappa}{6M}}\left[(\frac{1}{M^{2}}-\frac{3}{4\kappa^{2}})\sin(\frac{\kappa}{M})+
\frac{1}{M\kappa}\cos(\frac{\kappa}{M})\right]$.

The effective spring constant $K_{eff}$ along vertical direction, may be obtained by taking the sum of $K_{Y}$ and $K_{W}$ and is found to be function of Mach number M. $K_{eff}$ has been plotted as a function of M in Figure 7 where it shows almost no variation for $M>0.4$. As the normalized ion flow velocity is further decreased, it shows a jump towards positive value at around M=0.2 and then starts decreasing and exhibits a dip at M=0.125. Around this value of M, the spring constant becomes negative maximum indicating attractive wake potential to be a dominant interaction. The behaviour of $K_{eff}$ shows that the dominant interaction among the dust particles keep on changing from  Debye-H$\ddot{u}$ckel type to wake and the Debye-H$\ddot{u}$ckel again with the change in ion flow velocity. To understand the relation between the effective interaction and self-diffusion of dust grains we have further evaluated self-diffusion coefficient for this regime of ion flow velocity and the results are plotted in Figure 8. For lower values of M, we observe a very different behaviour of self-diffusion coefficient. As the mach number is varied from 0.2 to 0.125, self-diffusion coefficient gradually increases which is in agreement with the previous results. When M is further reduced self-diffusion coefficient starts decreasing again, thus exhibiting a kink corresponding to M=0.125. At this value of ion flow velocity, the wake potential becomes dominant and thus dislocates the particles from regular ordered position maintained due to repulsive Debye-H$\ddot{u}$ckel potential leading to increased diffusion of the dust grains. Below M=0.125, the effective spring constant starts increasing towards the positive value indicative a repulsive interaction dominant regime and during these values of M, self-diffusion coefficient decreases again. This is further supported by the plots of radial distribution function g(r) in Figure 9 and static structure factor S(k) in Figure 10.

\end{paragraph}
\begin{paragraph}{\textrm{IV.}}\textbf{CONCLUSIONS}

In this manuscript we have studied the self-diffusion coefficient of strongly coupled plasma in presence of ion streaming. Repulsive Debye-H$\ddot{u}$ckel and attractive wake potential have been used to perform Langevin dynamics simulation that gives the dynamics of dust particles. The simulation has been performed in subsonic regimes and corresponding self-diffusion coefficient has been calculated.
\begin{figure}[t]
  \centering
    \includegraphics[width=1.0\textwidth]{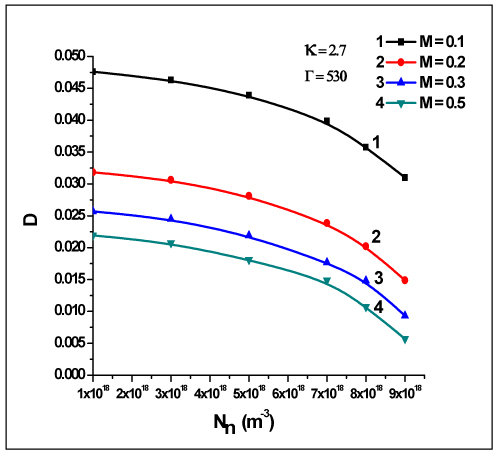}
    \caption[Plot of self-diffusion coefficient $D$ vs. neutral pressure $N_{n}$ for different $M$]{Plot of self-diffusion coefficient $D$ vs. neutral pressure $N_{n}$ for different Mach numbers at constant screening parameter $\kappa$=2.7 and Coulomb coupling parameter $\Gamma$=530. $D$ decrease with the increase in $N_{n}$ and $M$.}
\end{figure}
\begin{figure}[t]
  \centering
    \includegraphics[width=1.0\textwidth]{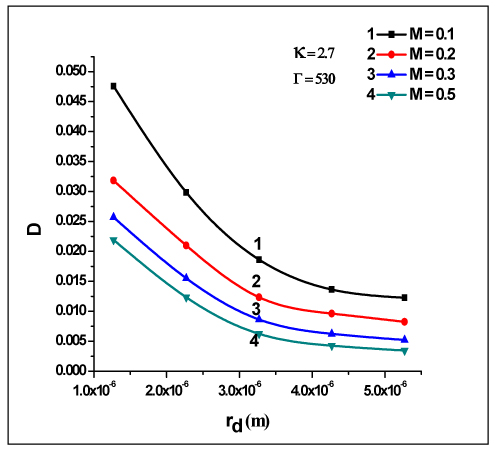}
    \caption[Plot of self-diffusion coefficient $D$ vs. dust radius $r_{d}$ for different $M$]{Plot of self-diffusion coefficient $D$ vs. dust radius $r_{d}$ for different Mach numbers at constant screening parameter $\kappa$=2.7 and Coulomb coupling parameter $\Gamma$=530. $D$ decrease with the increase in dust size and $M$.}
\end{figure}
\begin{figure}[t]
  \centering
    \includegraphics[width=1.0\textwidth]{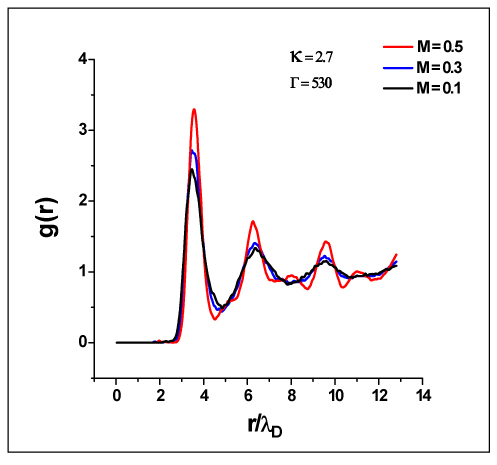}
    \caption[Plot of radial distribution function $g(r)$ as a function of $r$
for three Mach numbers]{Plot shows the radial distribution function $g(r)$ as a function of $r$
for three Mach numbers at constant screening parameter $\kappa$=2.7 and
Coulomb coupling parameter $\Gamma$=530. The red line corresponds to $M$=0.5,
the blue line corresponds to $M$=0.3 and the black line corresponds to
$M$=0.1. With the increase in $M$, peak height increase.}
\end{figure}
\begin{figure}[t]
  \centering
    \includegraphics[width=1.0\textwidth]{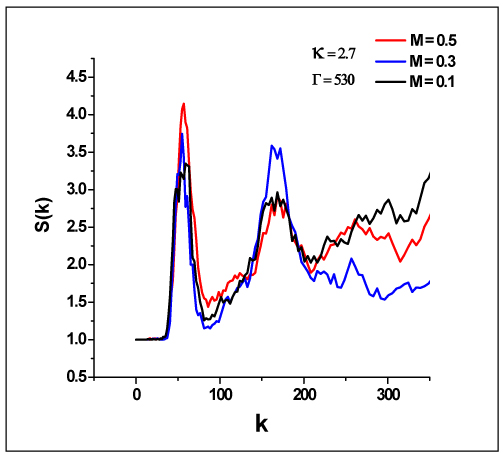}
  \caption[Plot of static structure factor $S(k)$ as a function of $k$ for
three Mach numbers]{Plot shows the static structure factor $S(k)$ as a function of $k$ for
three Mach numbers at constant screening parameter $\kappa$=2.7 and Coulomb
coupling parameter $\Gamma$=530. The red line corresponds to $M$=0.5, the blue
line corresponds to $M$=0.3 and the black line corresponds to $M$=0.1. Peak
height increase with the increase in $M$.}  
\end{figure}
\begin{figure}[t]
  \centering

    \includegraphics[width=1.0\textwidth=1.0]{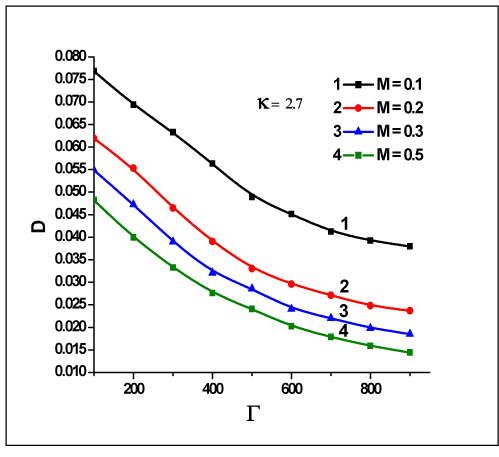}
    \caption[Plot of self-diffusion coefficient $D$ vs. Coulomb coupling
parameter $\Gamma$ for different Mach numbers]{Plot of self-diffusion coefficient $D$ vs. Coulomb coupling
parameter $\Gamma$ for different Mach numbers $M$=0.1, 0.2, 0.3 and 0.5 at
constant screening parameter $\kappa$=2.7 in the subsonic regime.}
\end{figure}

  \begin{figure}[t]
  \centering
    \includegraphics[width=1.0\textwidth]{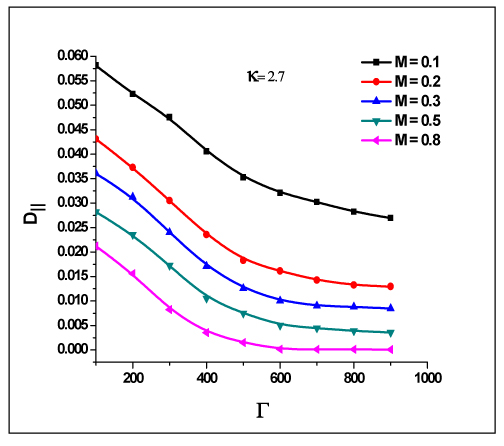}
\caption[Plot of self-diffusion coefficient $D_{\parallel}$ vs. Coulomb coupling parameter $\Gamma$]{Plot of self-diffusion coefficient $D_{\parallel}$ vs. Coulomb coupling parameter $\Gamma$ for different Mach numbers M=0.1, 0.2, 0.3, 0.5 and 0.8 at constant screening parameter $\kappa$=2.7.}
\end{figure}

\begin{figure}[t]
  \centering
    \includegraphics[width=1.0\textwidth]{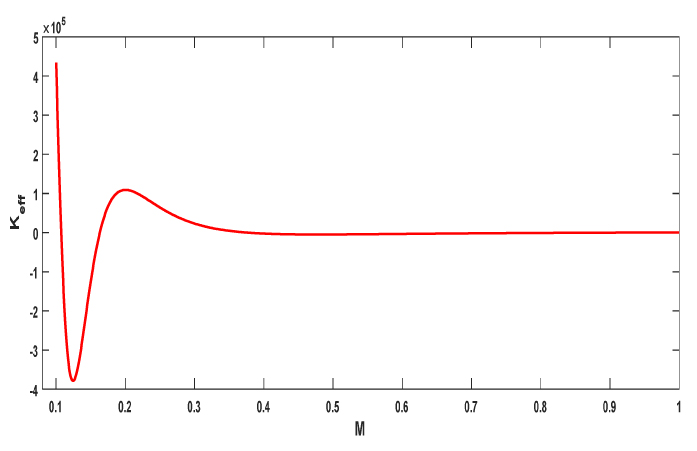}
\caption[Plot of effective spring constant $K_{eff}$ vs. Mach number $M$]{Effective spring constant $K_{eff}$ vs. Mach number $M$ for screening parameter $\kappa$=2.7 and Coulomb coupling parameter $\Gamma$=530.}
\end{figure}
\begin{figure}[t]
  \centering

    \includegraphics[width=1.0\textwidth]{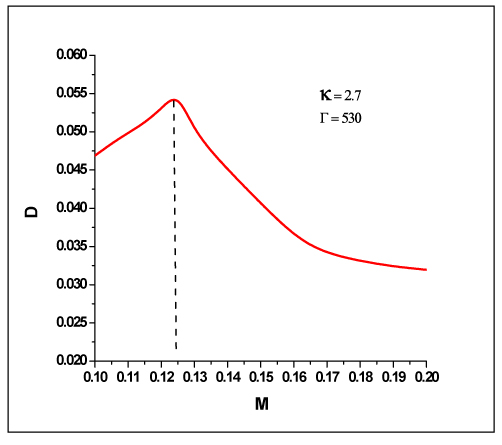}
    \caption[Plot of self-diffusion coefficient $D$ vs. Mach number $M$]{Plot of self-diffusion coefficient $D$ vs. Mach number $M$ at constant screening parameter $\kappa$=2.7 and Coulomb coupling parameter $\Gamma$=530.}
\end{figure}

  \begin{figure}[t]
  \centering
    \includegraphics[width=1.0\textwidth]{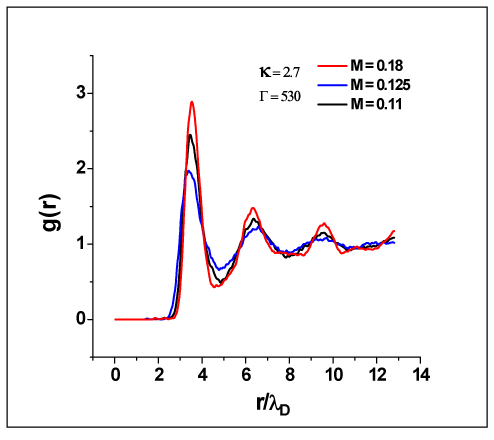}
\caption[Plot of radial distribution function $g(r)$ as a function of $r$
 for different Mach numbers]{Plot shows the radial distribution function $g(r)$ as a function of $r$
 for different Mach numbers at constant screening parameter $\kappa$=2.7 and
Coulomb coupling parameter $\Gamma$=530. The red line corresponds to $M$=0.18,
the blue line corresponds to $M$=0.125 and the black line corresponds to
$M$=0.11.}
\end{figure}
\begin{figure}[t]
  \centering
    \includegraphics[width=1.0\textwidth]{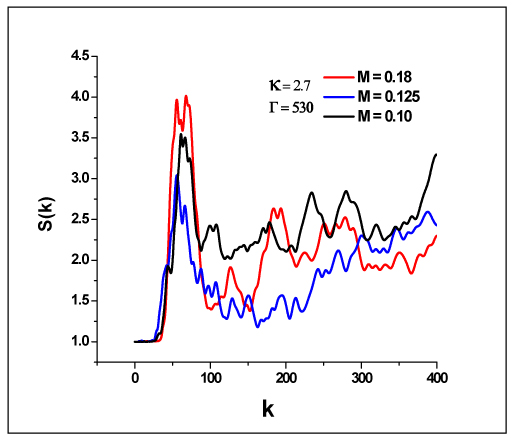}
    \caption[Plot of static structure factor $S(k)$ as a function of $k$ for
different Mach numbers]{Plot shows the static structure factor $S(k)$ as a function of $k$ for
different Mach numbers at constant screening parameter $\kappa$=2.7 and
Coulomb coupling parameter $\Gamma$=530. The red line corresponds to $M$=0.18,
the blue line corresponds to $M$=0.125 and the black line corresponds to
$M$=0.11.}

\end{figure}

\end{paragraph}

\begin{paragraph}{\textrm{V.}}\textbf{REFERENCES}
\newline
[1] M. Kroll, J. Schablinski, D. Block, and A. Piel, Phys. Plasmas \textbf{17}, 13702 (2010).
\newline
[2] O. Ishihara and S. V. Vladimirov, Phys. Plasmas \textbf{4}, 69 (1997).
\newline
[3] K. Takahashi, T. Oishi, K. Shimomai, Y. Hayashi,and S. Nishino, Phys. Rev. E \textbf{58}, 7805 (1998).
\newline
[4] D. S. Lemons, M. S. Murillo, W. Daughton, and D. Winske, Phys. Plasmas \textbf{7}, 2306 (2000).
\newline
[5] M. Nambu, S. V. Vladimirov, and P. K. Shukla, Phys. Lett. A \textbf{203}, 40 (1995).
\newline
[6] S. Bhattacharjee and N. Das, Phys. Plasmas \textbf{19}, 103707 (2012).
\newline
[7] A. K. Mukhopadhyay and J. Goree, Phys. Rev. E \textbf{90}, 013102 (2014).
\newline
[8]  H. Thomas, G. E. Morfill, V. Demmel, J. Goree, B. Feuerbacher, and D. M\"{o}hlmann, Phys. Rev. Lett. \textbf{73}, 652 (1994).
\newline
[9] S. Hamaguchi and  R. T. Farouki, D. H. E. Dubin, Phys. Rev. E \textbf{56}, 4671 (1997).
\newline
[10] X. H. Zheng and J. C. Earnshaw, Phys. Rev. Lett. \textbf{75}, 4214 (1995).
\newline
[11] W. K. Qi, Z. Wang, Y, Han, and Y. Chen, J. Chem. Phys. \textbf{133}, 234508 (2010).
\newline
[12] I. H. Hutchinson and C. B. Haakonsen, Phys. Plasmas \textbf{20}, 083701 (2013).
\newline
[13] P. Ludwig, W. J. Miloch, Hanno $K\ddot{a}hlert$, and M. Bonitz, New J. Phys. \textbf{14}, 053016 (2012).
\newline
[14] P. Bezbaruah and  N. Das, Phys. Plasmas \textbf{23}, 043701 (2016).
\newline
[15] S. A. Maiorov and B. A. Klumov, Bulletin of the Lebedev Physics Institute \textbf{40},(2013).
\newline
[16] D. C. Rapaport, \emph{The Art of Molecular Dynamics Simulation} (United Kingdom: Cambridge University press, 1995).
\newline
\end{paragraph}

\end{document}